# Detect adverse drug reactions for drug Pioglitazone

Yihui Liu[1,2]
[1]Institute of Intelligent Information Processing
Shandong Polytechnic University, China
Yihui_liu_2005@yahoo.co.uk

Uwe Aickelin[2]
[2]Department of Computer Science,
University of Nottingham, UK

*Abstract*—Adverse drug reaction (ADR) is widely concerned for public health issue. In this study we propose an original approach to detect the ADRs using feature matrix and feature selection. The experiments are carried out on the drug Pioglitazone. Major side effects for the drug are detected and better performance is achieved compared to other computerized methods. The detected ADRs are based on the computerized method, further investigation is needed.

## I. INTRODUCTION

Adverse drug reaction (ADR) is widely concerned for public health issue. ADRs are one of most common causes to withdraw some drugs from market [1]. Now two major methods for detecting ADRs are spontaneous reporting system (SRS) [2,3], and prescription event monitoring (PEM) [4,5]. The World Health Organization (WHO) defines a signal in pharmacovigilance as "any reported information on a possible causal relationship between an adverse event and a drug, the relationship being unknown or incompletely documented previously"[6]. For spontaneous reporting system, many machine learning methods are used to detect ADRs, such as Bayesian confidence propagation neural network (BCPNN) [7], decision support method [8], genetic algorithm [9], knowledge based approach [10], etc. One limitation is the reporting mechanism to submit ADR reports [8], which has serious underreporting and is not able to accurately quantify the corresponding risk. Another limitation is hard to detect ADRs with small number of occurrences of each drug-event association in the database.

In this paper we propose feature selection approach to detect ADRs from The Health Improvement Network (THIN) database. First feature matrix, which represents the medical events for the patients before and after taking drugs, is created by linking patients' prescriptions and corresponding medical events together. Then significant features are selected based on feature selection methods, comparing the feature matrix before patients take drugs with one after patients take drugs. Finally the significant ADRs can be detected from thousands of medical events based on corresponding features. Experiments are carried out on the drug Pioglitazone. Good performance is achieved.

## II. FEATURE MATRIX AND FEATURE SELECTION

### A. The Extraction of Feature Matrix

To detect the ADRs of drugs, first feature matrix is extracted from THIN database, which describes the medical events that patients occur before or after taking drugs. Then feature selection method of Student's t-test is performed to select the significant features from feature matrix containing thousands of medical events. Figure 1 shows the process to detect the ADRs using feature matrix. Feature matrix *A* describes the medical events for each patient during 60 days before they take drugs. Feature matrix *B* reflects the medical events during 60 days after patients take drugs. In order to reduce the effect of the small events, and save the computation time and space, we set 100 patients as a group. Matrix *X* and *Y* are feature matrix after patients are divided into groups.

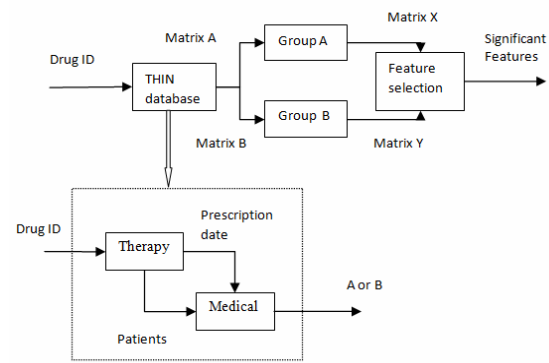

Figure 1. The process to detect ADRs. Matrix *A* and *B* are feature matrix before patients take drugs or after patients take drugs. The time period of observation is set to 60 days. Matrix *X* and *Y* are feature matrix after patients are divided into groups. We set 100 patients as one group.

### B. Medical Events and Readcodes

Medical events or symptoms are represented by medical codes or Readcodes. There are 103387 types of medical events in "Readcodes" database. The Read Codes used in general practice (GP), were invented and developed by Dr James Read in 1982. The NHS (National Health Service) has expanded the codes to cover all areas of clinical practice. The code is hierarchical from left to right or from level 1 to level 5. It means that it gives more detailed information from level 1 to level 5. Table 1 shows the medical symptoms based on Readcodes at level 3 and at level 5. 'Other soft tissue disorders' is general term using Readcodes at level 3. 'Foot pain', 'Heel pain', etc., give more details using Readcodes at level 5.

## C. Feature Selection Based on Student's t-test

Feature extraction and feature selection are widely used in biomedical data processing [11-18]. In our research we use Student's t-test [19] feature selection method to detect the significant ADRs from thousands of medical events. Student's t-test is a kind of statistical hypothesis test based on a normal distribution, and is used to measure the difference between two kinds of samples.

TABLE I. MEDICAL EVENTS BASED ON READCODES AT LEVEL 3 AND LEVEL 5.

|  | Level | Readcodes | Medical events |
|---|---|---|---|
| Muscle pain | Level 3 | N24..00 | Other soft tissue disorders |
|  | Level 5 | N245.16 | Leg pain |
|  |  | N245111 | Toe pain |
|  |  | N245.13 | Foot pain |
|  |  | N245700 | Shoulder pain |
|  |  | N245.15 | Heel pain |

## D. Other Parameters

The variable of ratio $R_1$ is defined to evaluate significant changes of the medical events, using ratio of the patient number after taking the drug to one before taking the drug. The variable $R_2$ represents the ratio of patient number after taking the drug to the number of whole population having one particular medical symptom.

The ratio variables $R_1$ and $R_2$ are defined as follows:

$$R_1 = \begin{cases} N_A / N_B & \text{if } N_B \neq 0; \\ N_A & \text{if } N_B = 0; \end{cases}$$

$$R_2 = N_A / N$$

where $N_B$ and $N_A$ represent the numbers of patients before or after they take drugs for having one particular medical event respectively. The variable $N$ represents the number of whole population who take drugs.

## III. EXPERIMENTS AND RESULTS

Pioglitazone [20] is a prescription drug of the class thiazolidinedione (TZD) with hypoglycemic (antihyperglycemic, antidiabetic) action to treat diabetes.

On July 30, 2007 an Advisory Committee of the Food and Drug Administration concluded that the use of rosiglitazone for the treatment of type 2 diabetes was associated with a greater risk of "myocardial ischemic events". On June 9, 2011 the French Agency for the Safety of Health Products decided to withdraw pioglitazone in regards to high risk of bladder cancer. On June 10, 2011 Germany's Federal Institute for Drugs and Medical Devices also advised doctors not to prescribe the medication until further investigation of the cancer risk had been conducted.

Pioglitazone [21] has the side effects: severe allergic reactions (rash; hives; itching; difficulty breathing; tightness in the chest; swelling of the mouth, face, lips, or tongue); blurred vision or other vision changes; swelling of the hands, ankles, or feet; symptoms of heart failure (shortness of breath; sudden, unexplained weight gain); symptoms of liver problems (dark urine; stomach pain; unexplained nausea, vomiting, or loss of appetite; yellowing of the skin or eyes); symptoms of low blood sugar (anxiety; chills, increased hunger, headache; increased dizziness or drowsiness; tremors); unusual bone pain; unusual tiredness or weakness, etc.

9093 patients from half of 475GP data in THIN database are taking Pioglitazone, and 11342 medical events are obtained based on Readcodes at level 1-5. After grouping them, 90x113412 feature matrix is obtained. For Readcodes at level 1-3, 90x2545 feature matrix is obtained.

Table 2 shows the top 30 detected results in ascending order of p value of Student's t-test, using Readcodes at level 1-5 and at level 1-3. The detected results are using p value less than 0.05, which represent the significant change after patients take the drug. Table 3 shows the results in descending order of the ratio of the number of patients after taking the drug to one before taking the drug. Table 4 shows the potential ADRs related to cancer for Pioglitazone based on p value of Student's t-test. Our detected results are consistent with the published side effects [20, 21]. Good performance is achieved.

MUTARA and HUNT [22，23] based on Unexpected Temporal Association Rule are proposed to signal unexpected and infrequent patterns characteristic of ADRs, using The Queensland Linked Data Set (QLDS). They indicated that "HUNT can reliably shortlist statistically significantly more ADRs than MUTARA". For the drug of Atorvastatin, HUNT detects 4 ADRs of 'urinary tract infection', 'stomach ulcer', 'diarrhoea', and 'bronchitis' from top 20 results based on 13712 patient records, and only obtains 30% accuracy [23]. The major side effects of Atorvastatin, which are 'muscle pain', 'muscle weakness', and other musculoskeletal events, are not detected. Table 5 shows the performance of HUNT and MUTARA.

TABLE V. THE ACCURACY OF HUNT AND MUTARA [23].

| Experimental settings | | Signalling accuracy | |
|---|---|---|---|
| Drug | Patients | HUNT | MUTARA |
| Atorvastatin | Older female | 0.30 | 0.20 |
|  | Older male | 0.30 | 0.20 |
|  | All patients (13712) | 0.30 | 0.20 |

## IV. CONCLUSIONS

In this study we propose a novel method to successfully detect the ADRs using feature matrix and feature selection. A feature matrix, which characterizes the medical events before patients take drugs or after patients take drugs, is created from THIN database. The feature selection method of Student's t-test is used to detect the significant features from thousands of medical events. The significant ADRs, which are corresponding to significant features, are detected. Experiments are performed on the drug Pioglitazone. Compared to other computerized method, our proposed method achieves good performance.

TABLE II. THE POTENTIAL TOP 30 ADRs FOR PIOGLITAZONE BASED ON P VALUE OF STUDENT'S T-TEST.

|  | Rank | Readcodes | Medical events | NB | NA | R1 | R2 |
|---|---|---|---|---|---|---|---|
| Level 1-5 | 1 | 1Z12.00 | Chronic kidney disease stage 3 | 133 | 623 | 4.68 | 6.85 |
| | 2 | 182..00 | Chest pain | 124 | 485 | 3.91 | 5.33 |
| | 3 | 1M10.00 | Knee pain | 141 | 546 | 3.87 | 6.00 |
| | 4 | N245.17 | Shoulder pain | 142 | 501 | 3.53 | 5.51 |
| | 5 | 1D14.00 | C/O: a rash | 98 | 403 | 4.11 | 4.43 |
| | 6 | N131.00 | Cervicalgia - pain in neck | 64 | 328 | 5.13 | 3.61 |
| | 7 | M03z000 | Cellulitis NOS | 61 | 277 | 4.54 | 3.05 |
| | 8 | 16C6.00 | Back pain without radiation NOS | 76 | 327 | 4.30 | 3.60 |
| | 9 | 1C9..00 | Sore throat symptom | 46 | 259 | 5.63 | 2.85 |
| | 10 | N143.00 | Sciatica | 37 | 221 | 5.97 | 2.43 |
| | 11 | 171..00 | Cough | 196 | 765 | 3.90 | 8.41 |
| | 12 | F420.00 | Diabetic retinopathy | 45 | 264 | 5.87 | 2.90 |
| | 13 | 1B5..11 | Dizziness symptom | 61 | 286 | 4.69 | 3.15 |
| | 14 | H01..00 | Acute sinusitis | 39 | 193 | 4.95 | 2.12 |
| | 15 | K190.00 | Urinary tract infection, site not specified | 65 | 268 | 4.12 | 2.95 |
| | 16 | D00..00 | Iron deficiency anaemias | 38 | 205 | 5.39 | 2.25 |
| | 17 | E227311 | Erectile dysfunction | 81 | 264 | 3.26 | 2.90 |
| | 18 | 1B8..00 | Eye symptoms | 30 | 169 | 5.63 | 1.86 |
| | 19 | H06z000 | Chest infection NOS | 130 | 479 | 3.68 | 5.27 |
| | 20 | 1739.00 | Shortness of breath | 54 | 278 | 5.15 | 3.06 |
| | 21 | F46..00 | Cataract | 31 | 154 | 4.97 | 1.69 |
| | 22 | N245200 | Pain in leg | 29 | 194 | 6.69 | 2.13 |
| | 23 | F4C0.00 | Acute conjunctivitis | 45 | 205 | 4.56 | 2.25 |
| | 24 | M15y100 | Intertrigo | 18 | 114 | 6.33 | 1.25 |
| | 25 | C380.00 | Obesity | 66 | 275 | 4.17 | 3.02 |
| | 26 | N142.11 | Low back pain | 83 | 357 | 4.30 | 3.93 |
| | 27 | 183..00 | Oedema | 46 | 273 | 5.93 | 3.00 |
| | 28 | N094K12 | Hip pain | 48 | 215 | 4.48 | 2.36 |
| | 29 | 1A55.00 | Dysuria | 50 | 197 | 3.94 | 2.17 |
| | 30 | C112.00 | Hypoglycaemia unspecified | 25 | 176 | 7.04 | 1.94 |
| Level 1-3 | 1 | 171..00 | Cough | 419 | 1501 | 3.58 | 16.51 |
| | 2 | N24..00 | Other soft tissue disorders | 445 | 1675 | 3.76 | 18.42 |
| | 3 | H06..00 | Acute bronchitis and bronchiolitis | 308 | 1124 | 3.65 | 12.36 |
| | 4 | 183..00 | Oedema | 133 | 694 | 5.22 | 7.63 |
| | 5 | N21..00 | Peripheral enthesopathies and allied syndromes | 150 | 652 | 4.35 | 7.17 |
| | 6 | 16C..00 | Backache symptom | 175 | 708 | 4.05 | 7.79 |
| | 7 | 1B1..00 | General nervous symptoms | 189 | 711 | 3.76 | 7.82 |
| | 8 | 173..00 | Breathlessness | 221 | 878 | 3.97 | 9.66 |
| | 9 | 1D1..00 | C/O: a general symptom | 202 | 771 | 3.82 | 8.48 |
| | 10 | H05..00 | Other acute upper respiratory infections | 142 | 634 | 4.46 | 6.97 |
| | 11 | 1M1..00 | Pain in lower limb | 171 | 675 | 3.95 | 7.42 |
| | 12 | 182..00 | Chest pain | 166 | 627 | 3.78 | 6.90 |
| | 13 | N14..00 | Other and unspecified back disorders | 139 | 650 | 4.68 | 7.15 |
| | 14 | 1C9..00 | Sore throat symptom | 51 | 323 | 6.33 | 3.55 |
| | 15 | N05..00 | Osteoarthritis and allied disorders | 109 | 488 | 4.48 | 5.37 |
| | 16 | M03..00 | Other cellulitis and abscess | 92 | 387 | 4.21 | 4.26 |
| | 17 | K19..00 | Other urethral and urinary tract disorders | 107 | 482 | 4.50 | 5.30 |
| | 18 | 1B8..00 | Eye symptoms | 82 | 396 | 4.83 | 4.35 |
| | 19 | F42..00 | Other retinal disorders | 195 | 850 | 4.36 | 9.35 |
| | 20 | N09..00 | Other and unspecified joint disorders | 188 | 744 | 3.96 | 8.18 |
| | 21 | N13..00 | Other cervical disorders | 66 | 352 | 5.33 | 3.87 |
| | 22 | F50..00 | Disorders of external ear | 114 | 449 | 3.94 | 4.94 |
| | 23 | F4C..00 | Disorders of conjunctiva | 71 | 334 | 4.70 | 3.67 |
| | 24 | 1Z1..00 | Chronic renal impairment | 220 | 838 | 3.81 | 9.22 |
| | 25 | G57..00 | Cardiac dysrhythmias | 30 | 165 | 5.50 | 1.81 |
| | 26 | 19F..00 | Diarrhoea symptoms | 202 | 524 | 2.59 | 5.76 |
| | 27 | 1C1..00 | Hearing symptoms | 42 | 240 | 5.71 | 2.64 |
| | 28 | 1B5..00 | Incoordination symptom | 74 | 349 | 4.72 | 3.84 |
| | 29 | D00..00 | Iron deficiency anaemias | 52 | 271 | 5.21 | 2.98 |
| | 30 | 168..00 | Tiredness symptom | 106 | 409 | 3.86 | 4.50 |

Variable $N_B$ and $N_A$ represent the numbers of patients before or after they take drugs for having one particular medical event. Variable $R_1$ represents the ratio of the numbers of patients after taking drugs to the numbers of patients before taking drugs. Variable $R_2$ represents the ratio of the numbers of patients after taking drugs to the number of the whole population.

TABLE III.   THE POTENTIAL TOP 30 ADRs FOR PIOGLITAZONE BASED ON DESCENDING ORDER OF R1 VALUE.

| | Rank | Readcodes | Medical events | NB | NA | R1 | R2 |
|---|---|---|---|---|---|---|---|
| Level 1-5 | 1 | 16J..00 | Swelling | 1 | 43 | 43.00 | 0.47 |
| | 2 | 1C84.00 | C/O - post nasal drip | 1 | 35 | 35.00 | 0.38 |
| | 3 | J15..00 | Gastritis and duodenitis | 1 | 29 | 29.00 | 0.32 |
| | 4 | 1B5..13 | Unsteady symptom | 1 | 28 | 28.00 | 0.31 |
| | 5 | A07y000 | Viral gastroenteritis | 1 | 28 | 28.00 | 0.31 |
| | 6 | G30..00 | Acute myocardial infarction | 1 | 27 | 27.00 | 0.30 |
| | 7 | 172..12 | Haemoptysis - symptom | 1 | 27 | 27.00 | 0.30 |
| | 8 | Fy03.11 | Obstructive sleep apnoea | 1 | 27 | 27.00 | 0.30 |
| | 9 | F4F2.00 | Epiphora | 1 | 23 | 23.00 | 0.25 |
| | 10 | SK17200 | Other ankle injury | 1 | 22 | 22.00 | 0.24 |
| | 11 | 1C6..00 | Nose bleed symptom | 0 | 20 | 20.00 | 0.22 |
| | 12 | N211.00 | Rotator cuff shoulder syndrome and allied disorders | 1 | 20 | 20.00 | 0.22 |
| | 13 | K197200 | Microscopic haematuria | 1 | 20 | 20.00 | 0.22 |
| | 14 | C294300 | Iron deficiency | 1 | 20 | 20.00 | 0.22 |
| | 15 | M101.12 | Seborrhoeic eczema | 1 | 19 | 19.00 | 0.21 |
| | 16 | AB0..12 | Ringworm | 0 | 18 | 18.00 | 0.20 |
| | 17 | F4C1411 | Allergic conjunctivitis | 2 | 36 | 18.00 | 0.40 |
| | 18 | G30..15 | MI - acute myocardial infarction | 0 | 17 | 17.00 | 0.19 |
| | 19 | G87..00 | Hypotension | 2 | 32 | 16.00 | 0.35 |
| | 20 | Eu01.00 | [X]Vascular dementia | 0 | 16 | 16.00 | 0.18 |
| | 21 | 1A24.00 | Stress incontinence | 0 | 16 | 16.00 | 0.18 |
| | 22 | 173C.12 | SOBOE | 2 | 32 | 16.00 | 0.35 |
| | 23 | N212200 | Subacromial impingement | 1 | 16 | 16.00 | 0.18 |
| | 24 | F425.00 | Degeneration of macula and posterior pole | 0 | 16 | 16.00 | 0.18 |
| | 25 | G573200 | Paroxysmal atrial fibrillation | 0 | 16 | 16.00 | 0.18 |
| | 26 | 1B87.00 | Has watering eyes | 0 | 16 | 16.00 | 0.18 |
| | 27 | F310.00 | Bell's (facial) palsy | 1 | 16 | 16.00 | 0.18 |
| | 28 | N245100 | Foot pain | 1 | 16 | 16.00 | 0.18 |
| | 29 | C10FJ00 | Insulin treated Type 2 diabetes mellitus | 4 | 61 | 15.25 | 0.67 |
| | 30 | 19FZ.11 | Diarrhoea & vomiting, symptom | 3 | 45 | 15.00 | 0.49 |
| | 1 | N12..00 | Intervertebral disc disorders | 1 | 24 | 24.00 | 0.26 |
| | 2 | C29..00 | Other nutritional deficiencies | 1 | 21 | 21.00 | 0.23 |
| | 3 | SD9..00 | Superficial injuries of multiple and unspecified sites | 1 | 19 | 19.00 | 0.21 |
| | 4 | B22..00 | Malignant neoplasm of trachea, bronchus and lung | 1 | 18 | 18.00 | 0.20 |
| | 5 | 1BD..00 | Harmful thoughts | 1 | 18 | 18.00 | 0.20 |
| | 6 | J07..00 | Salivary gland diseases | 1 | 16 | 16.00 | 0.18 |

| Level 1-3 | 7 | S33..00 | Fracture of tibia and fibula | 1 | 16 | 16.00 | 0.18 |
|---|---|---|---|---|---|---|---|
| | 8 | F31..00 | Facial nerve disorders | 1 | 16 | 16.00 | 0.18 |
| | 9 | C28..00 | Vitamin D deficiency | 1 | 16 | 16.00 | 0.18 |
| | 10 | 159..00 | H/O:gynaecological problem NOS | 1 | 15 | 15.00 | 0.16 |
| | 11 | 1C2..00 | Tinnitus symptoms | 1 | 15 | 15.00 | 0.16 |
| | 12 | 172..00 | Blood in sputum - haemoptysis | 2 | 29 | 14.50 | 0.32 |
| | 13 | N35..00 | Acquired deformities of toe | 1 | 14 | 14.00 | 0.15 |
| | 14 | 1BO..00 | Mood swings | 1 | 14 | 14.00 | 0.15 |
| | 15 | SL...15 | Overdose of drug | 0 | 13 | 13.00 | 0.14 |
| | 16 | Fy0..00 | Sleep disorders | 4 | 50 | 12.50 | 0.55 |
| | 17 | H2z..00 | Pneumonia or influenza NOS | 0 | 12 | 12.00 | 0.13 |
| | 18 | Eu5..00 | [X]Behav synd assoc with physiolgcl disturb + physical fctrs | 0 | 12 | 12.00 | 0.13 |
| | 19 | 1J6..00 | Suspected heart disease | 1 | 12 | 12.00 | 0.13 |
| | 20 | D0...00 | Deficiency anaemias | 1 | 12 | 12.00 | 0.13 |
| | 21 | SH...00 | Burns | 0 | 11 | 11.00 | 0.12 |
| | 22 | SN4..00 | Other external effect cause | 0 | 11 | 11.00 | 0.12 |
| | 23 | SD6..00 | Superficial injury of lower limb, excluding foot | 1 | 11 | 11.00 | 0.12 |
| | 24 | B49..00 | Malignant neoplasm of urinary bladder | 1 | 11 | 11.00 | 0.12 |
| | 25 | S89..00 | Other open wounds of other sites, excluding limbs | 1 | 11 | 11.00 | 0.12 |
| | 26 | TJC..00 | Adverse reaction to cardiovascular system drugs | 0 | 11 | 11.00 | 0.12 |
| | 27 | H1y..00 | Other specified diseases of upper respiratory tract | 4 | 43 | 10.75 | 0.47 |
| | 28 | F30..00 | Trigeminal nerve disorders | 2 | 21 | 10.50 | 0.23 |
| | 29 | SE3..00 | Contusion, upper limb | 2 | 20 | 10.00 | 0.22 |
| | 30 | N36..00 | Other acquired limb deformity | 0 | 10 | 10.00 | 0.11 |

TABLE IV. THE POTENTIAL ADRs RELATED TO CANCER FOR PIOGLITAZONE BASED ON P VALUE OF STUDENT'S T-TEST.

| Rank | Readcodes | Medical events | NB | NA | R1 | R2 |
|---|---|---|---|---|---|---|
| 1 | B76..00 | Benign neoplasm of skin | 15 | 73 | 4.87 | 0.80 |
| 2 | B22..00 | Malignant neoplasm of trachea, bronchus and lung | 1 | 18 | 18.00 | 0.20 |
| 3 | B33..00 | Other malignant neoplasm of skin | 17 | 48 | 2.82 | 0.53 |
| 4 | B34..00 | Malignant neoplasm of female breast | 3 | 22 | 7.33 | 0.24 |
| 5 | BB2..00 | [M]Papillary and squamous cell neoplasms | 4 | 21 | 5.25 | 0.23 |
| 6 | BB5..00 | [M]Adenomas and adenocarcinomas | 7 | 24 | 3.43 | 0.26 |
| 7 | B49..00 | Malignant neoplasm of urinary bladder | 1 | 11 | 11.00 | 0.12 |
| 8 | B46..00 | Malignant neoplasm of prostate | 4 | 22 | 5.50 | 0.24 |
| 9 | B8...00 | Carcinoma in situ | 2 | 13 | 6.50 | 0.14 |
| 10 | B14..00 | Malignant neoplasm of rectum, rectosigmoid junction and anus | 1 | 9 | 9.00 | 0.10 |
| 11 | B13..00 | Malignant neoplasm of colon | 3 | 14 | 4.67 | 0.15 |
| 12 | B59..00 | Malignant neoplasm of unspecified site | 0 | 5 | 5.00 | 0.05 |
| 13 | B83..00 | Carcinoma in situ of breast and genitourinary system | 1 | 7 | 7.00 | 0.08 |
| 14 | B71..00 | Benign neoplasm of other parts of digestive system | 3 | 12 | 4.00 | 0.13 |
| 15 | B51..00 | Malignant neoplasm of brain | 0 | 4 | 4.00 | 0.04 |
| 16 | B430200 | Malignant neoplasm of endometrium of corpus uteri | 0 | 3 | 3.00 | 0.03 |
| 17 | B80..00 | Carcinoma in situ of digestive organs | 0 | 3 | 3.00 | 0.03 |
| 18 | B17..00 | Malignant neoplasm of pancreas | 1 | 5 | 5.00 | 0.05 |
| 19 | B58..00 | Secondary malignant neoplasm of other specified sites | 1 | 5 | 5.00 | 0.05 |
| 20 | BBH..00 | [M]Myxomatous neoplasms | 0 | 2 | 2.00 | 0.02 |
| 21 | B21..00 | Malignant neoplasm of larynx | 0 | 2 | 2.00 | 0.02 |
| 22 | B11..00 | Malignant neoplasm of stomach | 0 | 2 | 2.00 | 0.02 |
| 23 | BB4..00 | [M]Transitional cell papillomas and carcinomas | 0 | 2 | 2.00 | 0.02 |
| 24 | BBG..00 | [M]Fibromatous neoplasms | 0 | 2 | 2.00 | 0.02 |
| 25 | B10..00 | Malignant neoplasm of oesophagus | 0 | 2 | 2.00 | 0.02 |
| 26 | B82..00 | Carcinoma in situ of skin | 0 | 2 | 2.00 | 0.02 |
| 27 | B53..00 | Malignant neoplasm of thyroid gland | 0 | 2 | 2.00 | 0.02 |
| 28 | BB1J.00 | [M]Small cell carcinoma NOS | 0 | 2 | 2.00 | 0.02 |
| 29 | B7E..00 | Benign neoplasms of eye | 1 | 4 | 4.00 | 0.04 |
| 30 | BB3..00 | [M]Basal cell neoplasms | 2 | 5 | 2.50 | 0.05 |